\documentclass[a4paper,11pt]{article}
\pdfoutput=1 
\usepackage{jheppub} 
\usepackage[T1]{fontenc} 
\usepackage{amsmath}
\usepackage{lineno}
\usepackage{siunitx}
\usepackage[noabbrev]{cleveref}
\usepackage{bm}
\newcommand{\dldx}{\text{d}\hat{\ell}/\text{d}x}
\newcommand{\dldxf}{\frac{\text{d}\hat{\ell}}{\text{d}x}}
\newcommand{\ltot}{\hat{\ell}_\text{tot}}
\newcommand{\lrad}{l_\text{rad}}

\crefname{equation}{eq.}{eqs.}

\title{\boldmath Probabilistic modeling of Cherenkov emission from particle showers}

\author[a]{Ian Crawshaw} 
\author[a]{Tianlu Yuan}
\author[a]{Emre Yildizci}
\author[a]{Lu Lu}
\author[b]{Anatoli Fedynitch}
\affiliation[a]{Dept. of Physics and Wisconsin IceCube Particle Astrophysics Center, University of Wisconsin{\textemdash}Madison, USA}
\affiliation[b]{Institute of Physics, Academia Sinica, Taiwan}

\emailAdd{icrawshaw@wisc.edu, tyuan@icecube.wisc.edu}

\abstract{
Subatomic particles can interact with target nuclei in matter or decay in flight, and an individual high-energy particle can induce a particle shower composed of numerous, lower-energy secondaries. These particle showers broadly exhibit universality across diverse media, including air, water, ice, and other materials, with their development governed by the Standard Model. Full Monte Carlo simulation of particle showers, where each secondary is individually tracked and propagated, can be a computational challenge to perform at scale. Experiments thus resort to parametrized approximations when efficient simulation becomes necessary. Here, we construct distributions of parameters capable of describing the Cherenkov light yield from particle showers in ice or water. Sampling from the distributions allows for a much improved description of event-to-event fluctuations, in amplitude and shape, along the shower axis. Including these effects is essential for a more accurate simulation of signal and background events in current and next-generation neutrino telescopes.
}

\begin{document}
\maketitle
\flushbottom

\section{Introduction}

Particle showers can be initiated by particle interactions in any medium and are central to many areas of physics. They are measured in collider and fixed-target high-energy physics (HEP) experiments, observed as jets from supermassive black holes, monitored in the atmosphere following cosmic-ray interactions with air, and anticipated from ultra-high-energy cosmic rays interacting with the cosmic microwave background. The secondary particles produced in these showers are critical observables for studying fundamental particle physics and probing the highest energy interactions in the Universe. The universality of particle showers through hadronic, electromagnetic (EM), and weak interactions allows for the development and sharing of tools to simplify the modeling of these interactions across the fields of particle physics and astrophysics.

Observation of Cherenkov radiation from charged secondary particles is the primary detection technique used in neutrino telescopes operating up to PeV energies. Typically, at energies above roughly \SI{10}{\giga \eV}, a neutrino interacts with the medium via $W^{\pm}$- or $Z^0$-mediated deep inelastic scattering (DIS), breaking the target nucleus and generating a hadronic shower along with a charged or neutral lepton of the same flavor. Each shower contains numerous charged particles that travel at a velocity $v$ exceeding the speed of light in the detection medium $c/n$, where $n$ is the phase index of refraction. These particles emit Cherenkov light, which propagates through an optically transparent medium such as ice or water, and can be detected by photosensors placed around the interaction vertex. To accurately interpret the detected signal and reconstruct properties of the incident neutrino, such as its flavor, energy, and direction, it is important to accurately model the Cherenkov emission arising from its interactions.

Two main characteristics of Cherenkov emission from a high-energy particle shower are the total amount of emission and the profile of this emission along the shower axis. The number of Cherenkov photons emitted by a charged particle travelling at a velocity $v$ greater than the critical velocity $c/n$, per unit length, $\ell$, and wavelength, $\lambda$, is given by the Frank-Tamm formula~\cite{Frank:1937fk}
    \begin{equation}
    \frac{\text{d}^2N}{\text{d}\ell\text{d}\lambda} = \frac{2\pi \alpha z^2}{\lambda^2} \left(1 - \frac{c^2}{v^2n^2} \right).
    \label{frank-tamm}
    \end{equation}
The weighted track length,
    \begin{equation}
        \hat{\ell}_i \equiv \ell_i \frac{n^2-c^2/v_i^2}{n^2 - 1}
        \label{eq:weight}
    \end{equation}
is defined as the distance a charged particle travelling at the speed of light would need to traverse to emit the same amount of Cherenkov light as particle $i$ travelling at velocity $v_i$ over a distance $\ell_i$. Summing over $i$, the total weighted track length of all charged particles, $\ltot$, can then be used as a measure of the total Cherenkov light emitted in a particle shower~\cite{Radel:2012ij,Raedel2012}. Furthermore, a gamma distribution with amplitude $\ltot$ can be used to parametrize the amount of Cherenkov light as a function of distance along the shower axis,
\begin{equation}
\label{eq:dldx}
    \dldxf = \ltot \frac{b^a}{\Gamma(a)}x^{a-1}e^{-bx},
\end{equation}
where $a$ and $b$ correspond to the usual shape and rate parameters of the gamma distribution, respectively.\footnote{In practice the emission is also of function of time, which can be approximated by $x/c$.} While a fully three-dimensional model would be a more realistic solution, our main goal here is to capture event-to-event differences in $\dldx$ and provide a more accurate model of $\ltot$.

This paper is structured as follows. \Cref{sec:sim} describes the  Monte Carlo (MC) simulations that form the basis for parametrizing $\dldx$ and $\ltot$. \Cref{sec:model} details the modeling of the simulated output and associated caveats. In \cref{sec:application}, we show how the model can be applied to high-energy neutrino interactions. Finally, \cref{sec:discuss} provides a summary of the main points and concludes.

\section{Simulation}
\label{sec:sim}

The first step is to produce a relatively large scale MC simulation of particle showers in ice. We use the FLUKA MC transport code~\cite{Ferrari:2005zk,Ballarini:2024isa} version \texttt{2025.1.0} to histogram the weighted track length of particle showers initiated by electrons, gammas, and several types of hadrons. Such MC simulations are accurate but come at a significant computational expense that increases with the initial particle energy. By generating a set of such simulations and performing a parametric model fit to each, we aim to obtain fast approximations to the overall Cherenkov photon yield and its shape.

\subsection{FLUKA configuration}
\label{sec:flukaconf}
FLUKA is a tool for simulating particle interactions with matter and their transport, capable of reaching PeV energies for hadrons with the DPMJET hadronic interaction model~\cite{Ballarini:2024isa}. Configuration involves specifying the properties of the initial particle, activating relevant physical effects, describing the physical medium, and selecting appropriate scoring methods and geometry.

To provide a primary-particle-specific parametrization, simulations were conducted with a fixed set of primary particles, chosen to represent the most common final-state hadrons in neutrino DIS interactions, as well as $e^-$ and $\gamma$. The simulated hadron primaries include $p,\ n,\ \pi^+,\ K^+,\ K^0_L,\ K^0_S,\ \Lambda^0,\ \Sigma^+,\ \Sigma^-,\ \Xi^0,\ \Xi^-$, and $\Omega^-$. Primary particles were individually injected into a medium of ice with density $\rho_0 = \SI{0.9216}{\g \per \cm^3}$, corresponding to that at the center of IceCube~\cite{IceCube:2013llx}. The initial kinetic energies of hadrons ($e^-$ and $\gamma$) were set to 51 (61) logarithmically spaced values from \SI{10}{\giga \eV} (\SI{1}{\giga \eV}) to \SI{1}{\peta \eV}, each with \si{10000} repeated injections of a single primary particle. Each injection was assigned a specified random seed to ensure reproducibility and enable further investigation. Besides muons, all secondary particles were propagated and allowed to interact until they fell below Cherenkov-threshold energies, corresponding to \SI{264}{\kilo \eV} for electrons, positrons, and photons. Default thresholds were in place for hadrons (e.g., \SI{100}{\kilo \eV} for charged hadrons). Any muons produced were stopped immediately by setting its threshold to an impossibly high energy. These secondary muons are most commonly produced in decays of hadrons at lower energies, and a separate treatment is typically employed~\cite{Panknin:2009}.

The FLUKA \texttt{DEFAULTS} were set to \texttt{PRECISIOn} to turn on transport of electrons, positrons and photons and enable more precise treatment for a variety of physics processes~\cite{Ballarini:2024isa}. Notable additional effects that were activated via the \texttt{PHYSICS} command include a new evaporation model with heavy fragment evaporation, coalescence, ion splitting into nucleons, photonuclear and electronuclear interactions, and full transportation of all light and heavy ions. These settings were chosen following guidance from the FLUKA developers, and are listed in detail in \cref{sec:appendix-fluka}.

The geometry for the medium was defined as a single block of ice of dimension $\SI{20}{\m} \times \SI{20}{\m} \times \SI{50.5}{\m}$. The last dimension corresponds to the directional or longitudinal axis of the primary particle, and the first two form the transverse plane. The primary particle is injected into the ice at the origin, which lies at the center of the transverse plane and \SI{0.5}{\m} in from one end of the block. All particles are terminated upon reaching the block's boundary by surrounding the material with FLUKA's predefined \texttt{BLCKHOLE} medium. As the primary goal of this study is to parametrize $\dldx$ along the longitudinal axis, a volume-based scoring option was selected. The medium was thus divided into \si{500} equally sized, one-dimensional bins ranging from \SIrange{0}{50}{\m}. Within each bin, the total Frank-Tamm corrected charged particle track length was calculated using \cref{eq:weight}, assuming $n=1.33$. As described in ref.~\cite{Radel:2012ij}, the use of the equivalent length allows for $\ltot$ to be rescaled to a different density and refractive index. Finally, it is more natural to parametrize $\dldx$ in units of $\lrad = (\SI{36.08}{\g \per \cm^2}) / \rho_0$, so that its shape can be easily rescaled for different densities.

\subsection{Characteristics of FLUKA results}
\label{sec:flukacomp}

\begin{figure}[tbp]
\includegraphics[width=0.51\linewidth]{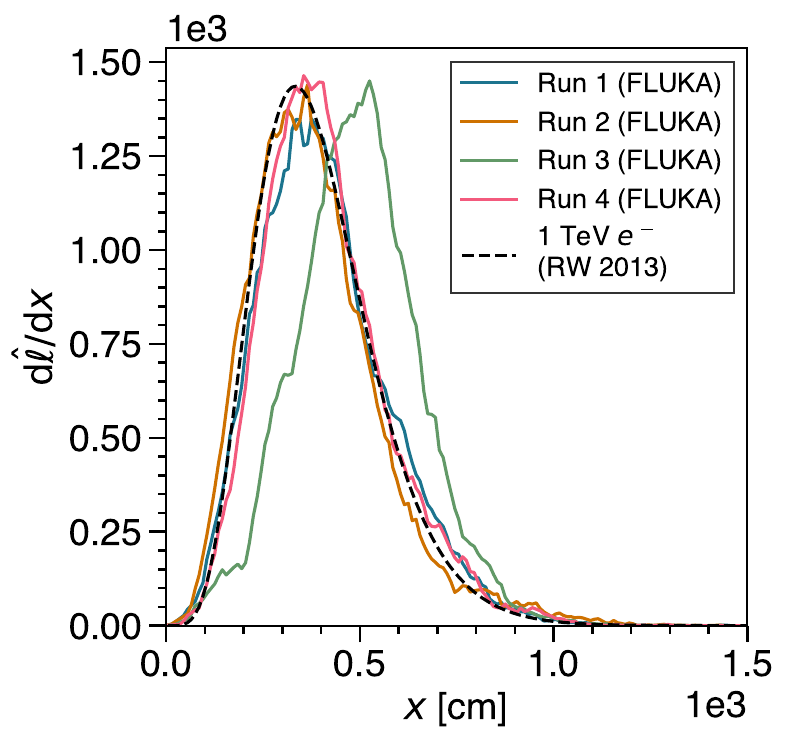}
\includegraphics[width=0.48\linewidth]{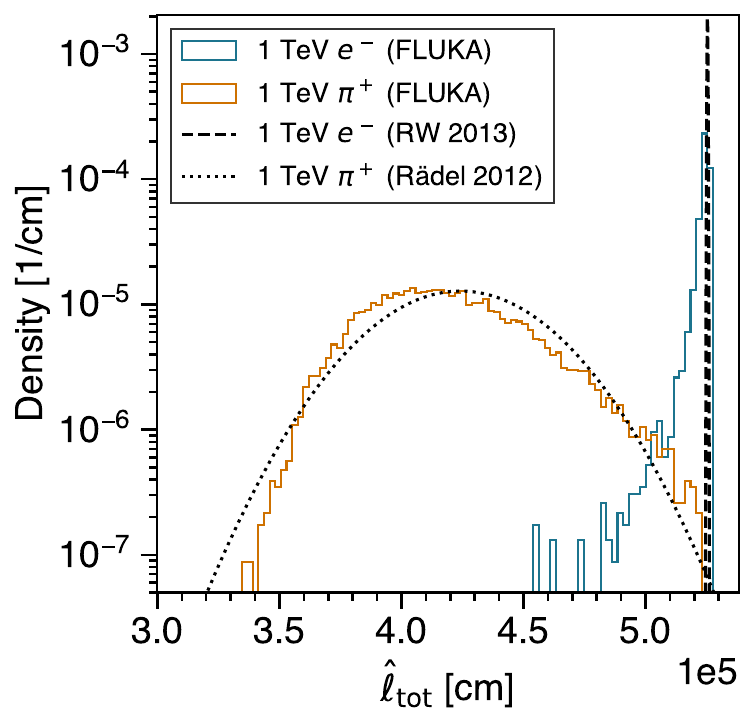}
\caption{The left panel shows four randomly chosen $\dldx$ distributions for a \SI{1}{\tera \eV} $e^-$ from FLUKA in solid, colored lines. The parametrization for $\dldx$ from ref.~\cite{Radel:2012ij} is shown as the dashed, black curve. The right panel shows the $\ltot$ distribution for a \SI{1}{\tera \eV} $e^-$ ($\pi^+$) from FLUKA in blue (orange). The corresponding $\ltot$ distribution as parametrized in refs.~\cite{Raedel2012,Radel:2012ij} is shown as a dashed (dotted) black curves. Note that the parametrized $\ltot$ curve for the \SI{1}{\tera \eV} $e^-$ has a very narrow width, and its mean is assumed for the $\dldx$ parametrization in the left panel.
\label{fig:example_showers}}
\end{figure}
While the average of many particle showers initiated by the same primary at a particular energy may be well approximated by a single gamma distribution~\cite{Radel:2012ij}, each individual shower exhibits deviations from the average. The left panel of \cref{fig:example_showers} shows four example $\dldx$ distributions obtained from independent FLUKA runs, in which a \SI{1}{\tera \eV} $e^-$ was injected, as solid, colored lines. For comparison, the average $\dldx$ distribution, parametrized in ref.~\cite{Radel:2012ij}, is shown as the dashed, black curve for the same primary. The four FLUKA runs were randomly chosen in this example, and such a comparison already demonstrates how the shower-to-shower $\dldx$ deviates from the mean, in both shape and amplitude.

The right panel of \cref{fig:example_showers} shows the $\ltot$ distribution for a \SI{1}{\tera \eV} $e^-$ ($\pi^+$) from FLUKA in blue (orange). For each run, $\ltot$ is simply the integral of $\dldx$ over $x$, and the panel thus shows a distribution over all runs. Correspondingly, the $\ltot$ distributions as parametrized in ref.~\cite{Raedel2012,Radel:2012ij} are shown as dashed (dotted) black curves. Both are Gaussians, with the $\ltot$ from $e^-$ exhibiting a much smaller variance than the corresponding distribution from FLUKA. The observed difference for $e^-$ also exists for $\gamma$-initiated showers, and is attributable to enabling photonuclear and electronuclear effects in FLUKA~\cite{Cerutti:2017tac}, which were not enabled for the Geant4-based parameterizations~\cite{Radel:2012ij}. The physics can be broadly understood as the emission of nucleons due to excitation by MeV electrons, which translates into a decrease in the number of particles above Cherenkov threshold.

\begin{figure}[tbp]
    \includegraphics[width=0.5\linewidth]{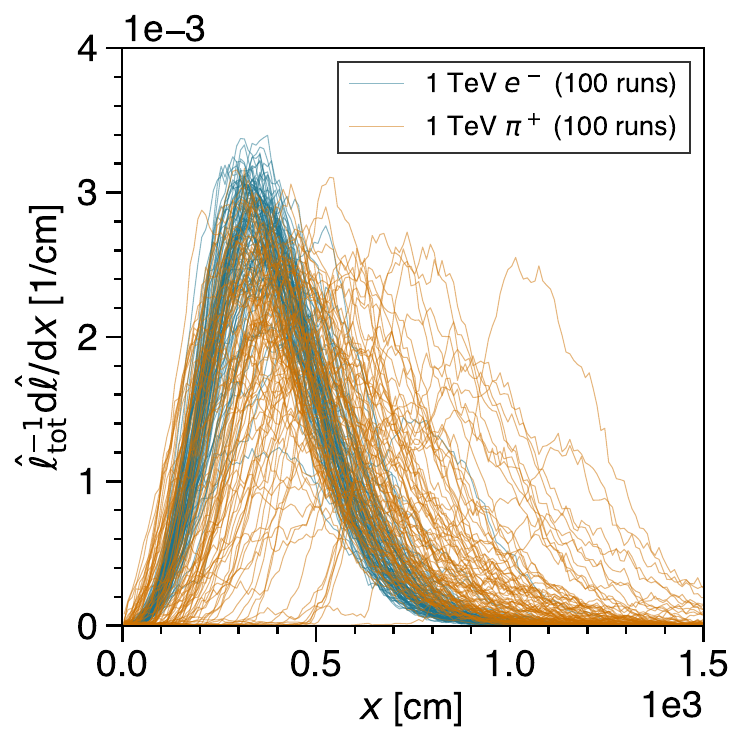}
    \includegraphics[width=0.51\linewidth]{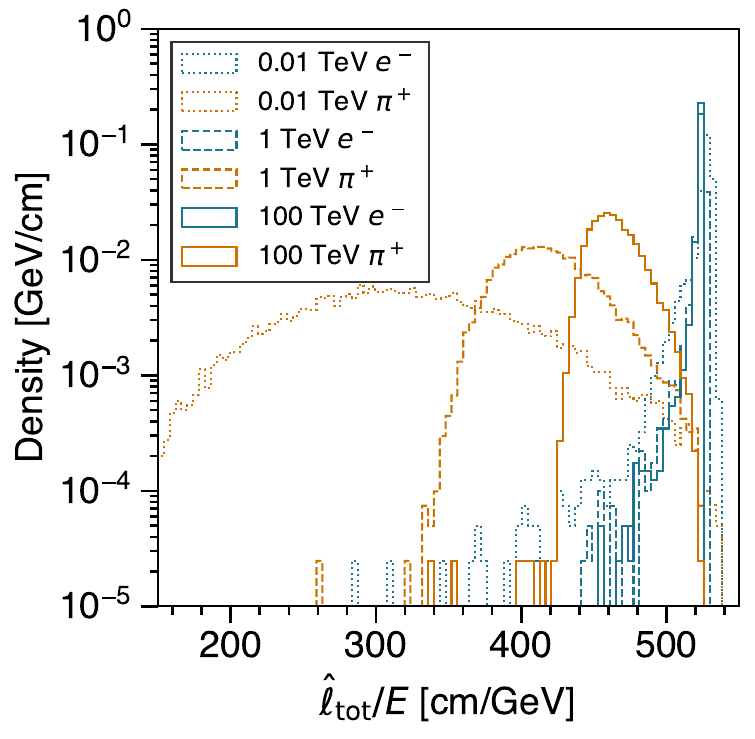}
    \caption{The left panel shows 100 $\ltot^{-1} \dldx$ distributions resulting from a \SI{1}{\tera \eV} $e^-$ (blue) and $\pi^+$ (orange), chosen at random. The right panel shows the distribution of $\ltot /E$ for $e^-$ (blue) and $\pi^+$ (orange) at three different energies as indicated in the legend. As is evident, $e^-$-initiated showers exhibits lower shower-to-shower fluctuations in both shape and amplitude compared to hadronic showers.}
    \label{fig:emvspi}
\end{figure}
A more in-depth visualization of FLUKA $\ltot^{-1} \dldx$ distributions can be found in the left panel of \cref{fig:emvspi}, which consists of 100 randomly selected profiles initiated by a \SI{1}{\tera \eV} $e^-$ ($\pi^+$) in blue (orange). We see that the blue lines from $e^-$ cluster more tightly together. In contrast, the distributions from a $\pi^+$ initiated shower exhibit a larger spread in shape, with a few peaking at upwards of \SI{10}{\m} from the injection point. More generally, hadron-initiated showers will exhibit larger variance in both $\ltot^{-1} \dldx$ and $\ltot$ than EM ($e^-$- or $\gamma$-initiated) showers~\cite{Raedel2012}. A comparison of the $\ltot/E$ distributions for $e^-$ (blue) and $\pi^+$ (orange) at energies of \SI{0.01}{\tera \eV}, \SI{1}{\tera \eV}, and \SI{100}{\tera \eV} is given in the right panel of \cref{fig:emvspi}. For $e^-$, the mean $\ltot$ grows linearly with its energy. For $\pi^+$, the $\ltot$ distribution has a much larger variance than its EM counterpart and deviates somewhat from linearity.

\begin{figure}[tbp]
    \centering
    \includegraphics[width=0.49\linewidth]{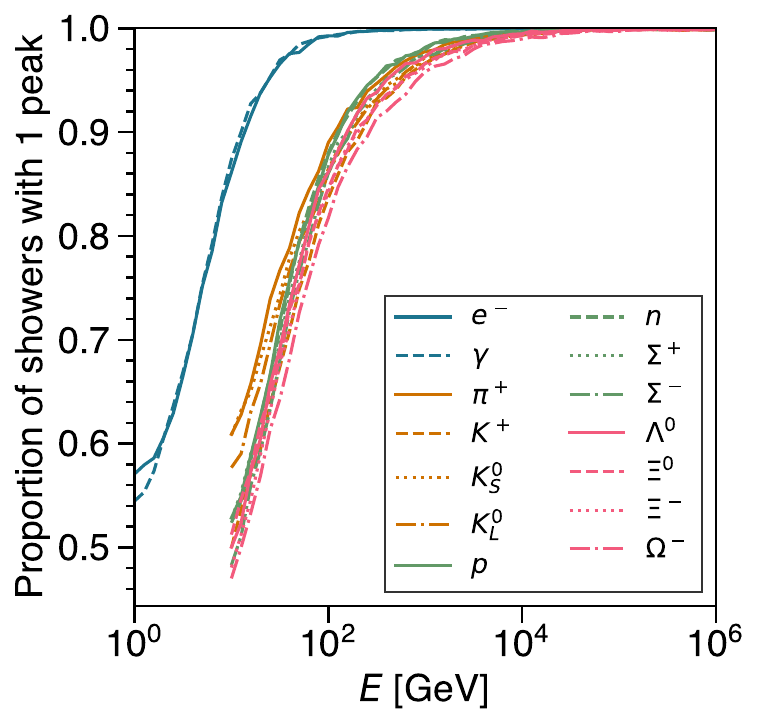}
    \includegraphics[width=0.49\linewidth]{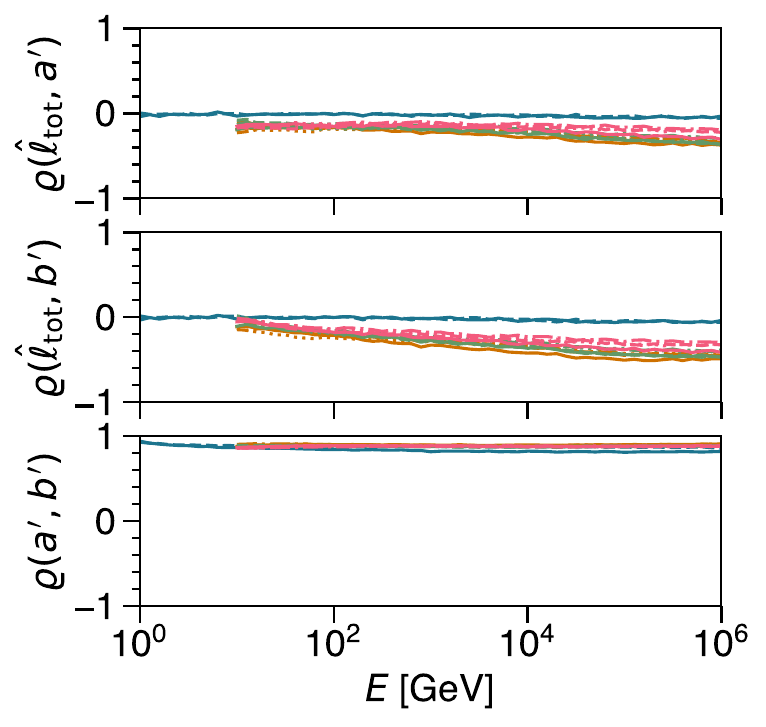}
    \caption{The left panel shows fraction of simulation runs resulting in single-peak shower profiles for different primary particles, plotted as a function of its initial kinetic energy. Electromagnetic showers initiated by $e^-$ and $\gamma$ are shown in blue, mesons in orange, and baryons in green and red. All trend towards one as energy increases, but exhibit differences at lower energies between EM and hadronic showers, with smaller differences visible between the mesons and baryons. The right panel shows Spearman's rank coefficient between $\ltot$, $a'$ and $b'$ (c.f.~\cref{eq:ab_transforms}), also as a function of energy. The same line style and color scheme, corresponding to different primary particles, is used as the left panel. Values of $\varrho(a', b')$ lie close to one, indicating strong correlation. Values of $\varrho(\ltot, \{a', b'\})$ for $e^-$ and $\gamma$ (blue) lie near zero, while hadronic shower amplitudes become more anticorrelated with $a'$ and $b'$ as energy increases. Note $e^-$ and $\gamma$ were simulated down to \SI{1}{\giga \eV}; hadrons \SI{10}{\giga \eV}.}
    \label{fig:statistics}
\end{figure}
As a summary statistic for the shower shape we computed the number of peaks for each $\dldx$. Peaks were identified by first smoothing shower profiles with a rolling average, and then identifying local maxima with prominence above 0.004 and width above \SI{30}{\cm}. \Cref{fig:statistics} (left) shows the fraction of simulated FLUKA showers found with only one peak as a function of the initial energy of the primary particle. Primary particles are classified by color into three groups, with $e^-$ and $\gamma$ in blue, mesons in orange, and baryons in green and red. Two trends are visible and worth highlighting, the first being the obvious difference between EM (blue) and hadronic showers (other colors). The second is the subdominant, but still visible, differences between the light mesons that were simulated (orange), in contrast to the EM and baryonic primaries, which cluster more closely together. The meson differences are likely due their varying branching ratios and lifetimes; their decays can impact how the first interaction proceeds.

Finally, we calculated Spearman's rank coefficient, $\varrho$, to assess the correlation between $\ltot$ and the shape and rate parameters of the gamma distribution. For each simulated shower, we fit $a$ and $b$ of \cref{eq:dldx} to $\ltot^{-1} \dldx$ using non-linear least squares regression,\footnote{As in ref.~\cite{Radel:2012ij} $x$ is in units of $\lrad$ and $b$ unitless, meaning the shape of the physical gamma distribution can be rescaled to varying densities.}. The rank coefficient results are shown in the right panels of \cref{fig:statistics}, with $\varrho(\ltot, a')$ (top), $\varrho(\ltot, b')$ (middle), and $\varrho(a', b')$ (bottom), where $a'$ and $b'$ are defined by
\begin{equation}
    \label{eq:ab_transforms}
    a' \equiv \frac{1}{\sqrt{a}}, \quad b' \equiv \frac{1}{1 + b^2}.
\end{equation}
The rationale for such a transformation will be discussed in \cref{sec:shape}. The same line and color style is used as in the left panel, corresponding to different primary particles. Values of $\varrho(a', b')$ lie close to one for all primaries, indicating strong correlation. In contrast, $\varrho(\ltot, \{a', b'\})$ for EM showers lie near zero, indicating a lack of correlation, while those for hadron primaries exhibit some anticorrelation as energy increases. For hadrons below roughly \SI{10}{\giga \eV}, the primary particle loses energy via ionization and excitation of electrons. Decays also occur with increasing probability for the mesons and heavier baryons. The resulting Cherenkov yield from these dominant, stochastic processes are difficult to describe using the parametrization in \cref{sec:model}, and to be safe a minimum energy of \SI{100}{\giga \eV} (\SI{10}{\giga \eV}) is recommended for hadrons (EM), though the support extends down from those values by an order of magnitude. At these energies, fully MC-based approaches are recommended instead, especially as the computational needs are also lower.

It is clear from this discussion that neither the shape nor scale of individual showers is represented well by the average shower. A more accurate model of particle showers must, to some extent, account for these fluctuations. As will be shown in \cref{sec:model}, parametric probability distributions offer a simple solution to model the $\ltot$ distribution, as well as $\ltot^{-1} \dldx$, from individual showers.

\section{Model}
\label{sec:model}

While the model from ref.~\cite{Radel:2012ij,Raedel2012} is unable to describe individual particle showers, each individual shape still largely resembles a gamma distribution. The parameters of this distribution can vary quite dramatically from shower to shower. We thus parametrize each shower with equation \cref{eq:dldx}, and seek a model that can generate the parameters $\ltot$, $a$, and $b$ as random variables. The $\ltot$ parameter controls the amplitude of the shower, while $a$ and $b$ control its shape. As shown in the right panel of \cref{fig:statistics}, $\ltot$ possesses only a loose correlation with the other two parameters. Thus, a simplifying assumption was made to treat $\ltot$ independently from the other two parameters. A more complete treatment would sample showers from the full three dimension space of $(a, b, \ltot)$ as a function of primary energy, but introduces extra complexity and computational requirements. Since shape variations are dominant, a fast and functional model for particle showers can be obtained by constructing probability distributions over the factorized space of $(a,b)$ and $\ltot$.

For hadron primaries, outliers were removed using two simple cuts. In order to avoid pulls from rare decays, in which a large fraction of primary energy can become invisible, at each energy simulated, those events with $\ltot$ falling into the lowest \SI{0.5}{\%} quantile were removed. Further, the Wasserstein distance between the fitted and simulated shapes was calculated. Events with distances in the highest $\SI{0.2}{\%}$ quantile, corresponding to the worst agreement, were removed. A more in-depth discussion of the Wasserstein distance will be given in \cref{sec:accuracy}.

\subsection{Shape variations with basis splines}
\label{sec:shape}

To capture variations in $(a, b)$, we first construct their probability distributions based on simulation data, conditioned on the energy of the initial particle, then sample from those distributions. The degree to which each profile can be represented by a gamma distribution varies, but it is an accurate approximation over a wide range of energies. As the primary particle energy decreases to GeVs, the resulting $\ltot^{-1} \dldx$ can deviate from the gamma distribution. Stronger deviations occur when showers have more than one peak, but \cref{fig:statistics} shows that this proportion becomes largely insignificant at energies above \SI{1}{\tera \eV}.

The range of valid values for the fitted $a$ and $b$ parameters are $(1,\infty)$ and $(0,\infty)$ respectively. Due to added computational difficulty sampling from a multivariate distribution with unbounded support, we model the joint probability density function (PDF), $f(a', b';E)$, in terms of \cref{eq:ab_transforms} for a given primary energy, thus restricting its support to lie on the unit square. In searching for transformations we wanted functions which maximized both computational simplicity and uniformity across the unit square. At each energy level we wanted to maximize resolution by spreading the data across the interval as evenly as possible. Both transforms are monotonic and so do not affect the rank correlations of the parameters. As $a'$ and $b'$ are highly correlated, a multivariate PDF is unavoidable.

\begin{figure}[tbp]
    \includegraphics[width=0.5\linewidth]{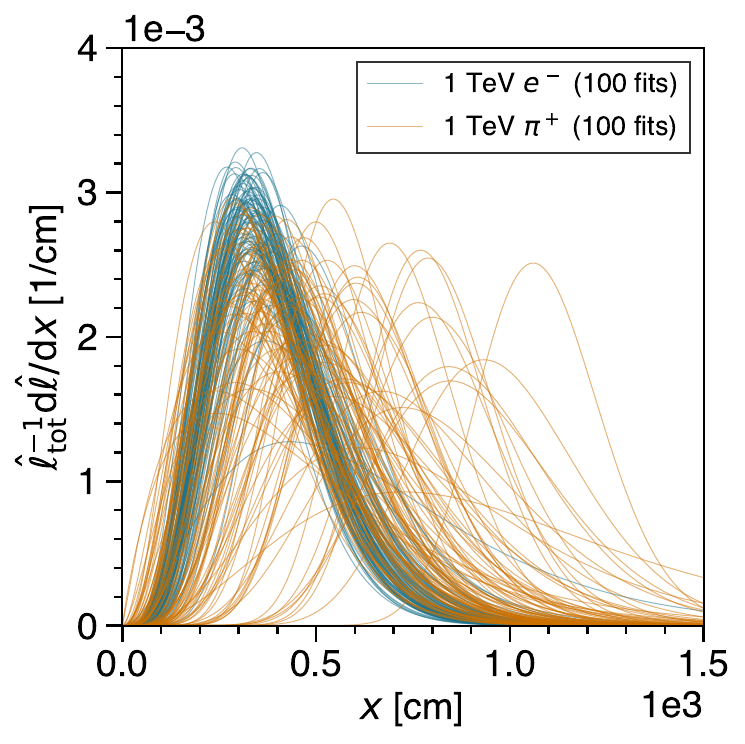}
    \includegraphics[width=0.51\linewidth]{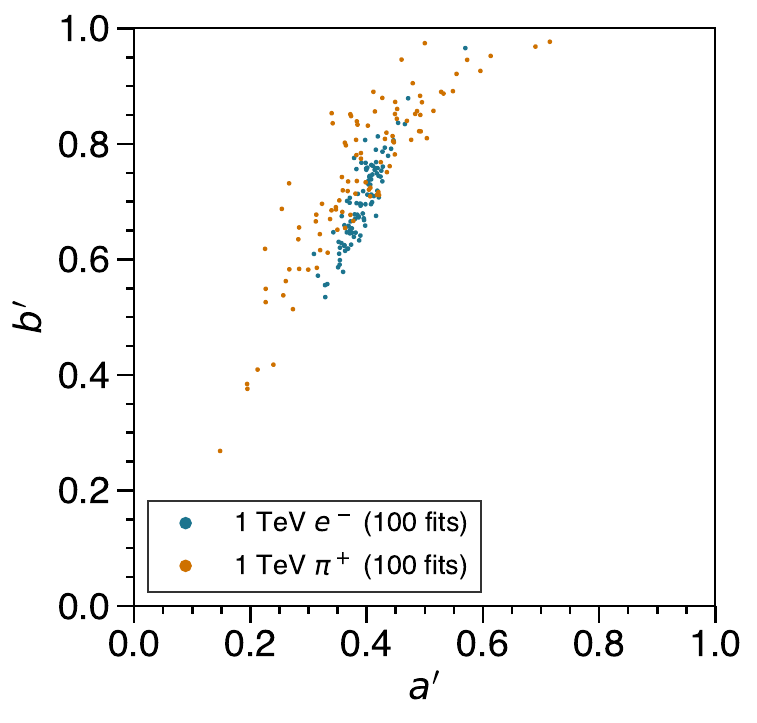}
    \caption{The left panel shows fitted $\ltot^{-1} \dldx$ distributions for the same set of simulations shown in the left panel of \cref{fig:emvspi}. Each shape is modeled by a two-parameter gamma distribution, which captures the per-shower longitudinal fluctuations. The right panel shows the fitted parameters $(a', b')$ for the distributions in the left panel, after applying \cref{eq:ab_transforms}. A clear distinction between $e^-$ and $\pi^+$ is visible with the denser $e^-$ cluster corresponding to its higher degree of similarity across different simulation runs.}
    \label{fig:emvspifits}
\end{figure}
The left panel of \cref{fig:emvspifits} shows fitted shapes for the same 100 simulation runs as the left panel of \cref{fig:emvspi}. Although much smoother than the exact MC profiles, a general resemblance in location and shape is evident. The right panel of \cref{fig:emvspifits} shows a scatter of the corresponding fit parameters $(a', b')$, after applying \cref{eq:ab_transforms}. The different clustering of $e^-$ and $\pi^+$ is indicative of the variations seen in EM and hadronic showers. The right panel highlights a subset of the data that is then binned and fitted, as described below.

\begin{figure}
    \centering
    \includegraphics[width=\linewidth]{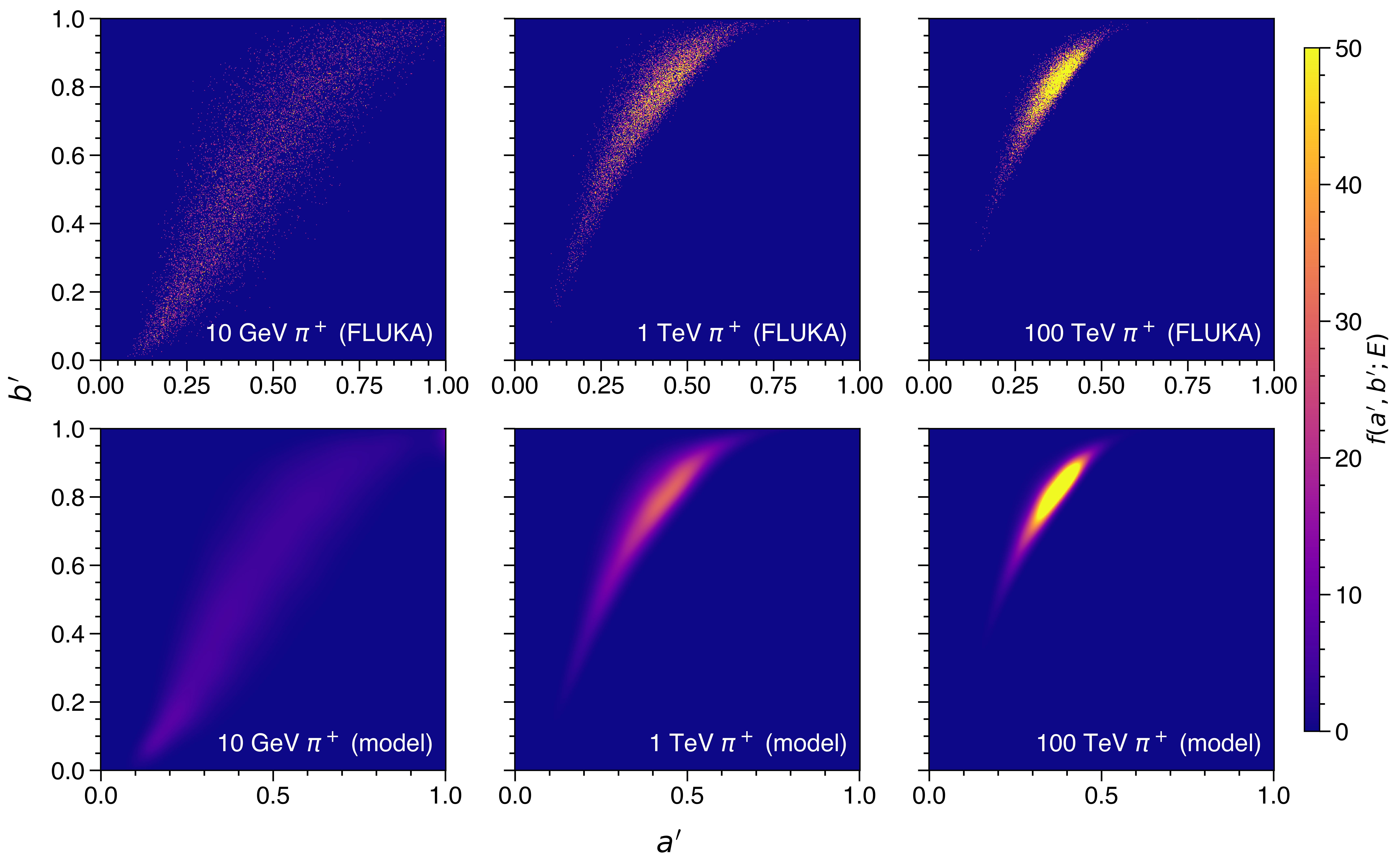}
    \caption{The top row shows normalized histograms of $(a', b')$ from FLUKA $\pi^+$ simulations at three different energies, \SI{10}{\giga \eV} (left), \SI{1}{\tera \eV} (middle) and \SI{100}{\tera \eV} (right), binned as described in the text. The bottom row shows the probability density function $f(a', b' ; E)$ as generated from our model, for the same primary and energies. The same color scale is used for all six panels. Furthermore, the middle panels can be compared to the orange scatter in the left panel of \cref{fig:emvspifits}.}
    \label{fig:ab_dists}
\end{figure}

At each energy, the samples of $(a', b')$ obtained from FLUKA MC were aggregated as a two-dimensional histogram with 450 by 450 equally sized bins over the unit square. Examples for $\pi^+$ at three different energies are shown in the top row of \cref{fig:ab_dists}. We use this histogram data to estimate $f(a', b';E)$ as a function of energy. The PDF is modeled with a tensor product of exponentiated, penalized B-splines following the procedure described by Eilers and Marx~\cite{10.1214/ss/1038425655}. The method is a generalized linear model (GLM), with the natural logarithm as the link function to a linear estimator. This link function ensures that the PDF is nonnegative. Error in each histogram bin is treated as being drawn from a Poisson distribution. The predictor, $\eta(a',b',E) = \ln f(a', b'; E)$ is a linear combination of basis splines, which in this case are chosen to be degree three. The PDF is then given by $\exp(\eta)$. If $B_{a',i}(a')$, $B_{b',j}(b')$, and $B_{E,k}(E)$ denote the $i$th, $j$th, and $k$th basis splines in each of the three dimensions, then a single basis element for our three-dimensional spline is given by their product. Our model is thus represented by the parameter array $\bm{\theta}$, such that $\eta(a',b',E) = \sum_{i,j,k}\theta_{ijk}B_{a',i}(a')B_{b',j}(b')B_{E,k}(E)$.

This method works best with finer histogram binning, but if binning is too high the time and space complexity of the fitting procedure can become intractable. A similar balance must be struck in the number of basis splines used to model the density function. We denote this number for the $a'$, $b'$ and $E$ dimensions as $c_{a'}$, $c_{b'}$, and $c_E$, respectively. More basis splines can resolve finer detail in the distribution, but can become computationally intensive. For the model discussed in this paper, $c_{a'} = c_{b'} = 17$ and for hadron primaries $c_{E} = 8$, otherwise $c_E = 9$. Knots locations were set equally spaced apart in $a'$, $b'$ and $\log_{10}(E)$, across the full extent of each dimension.

In order to limit over-fitting the histogram we impose a penalty term when fitting the spline. In traditional GLMs the objective is to maximize the log-likelihood function, $\ell(\bm{\theta})$. In a penalized model, a smoothness dependent term is additionally subtracted from the log-likelihood. In our model we maximize $\ell(\bm{\theta}) - \frac{\lambda_{a'}}{2}D_{a'}(\bm{\theta}) - \frac{\lambda_{b'}}{2}D_{b'}(\bm{\theta}) - \frac{\lambda_E}{2}D_E(\bm{\theta})$. Each $\lambda_i$ is a parameter chosen to scale the smoothing strength in the $i$th dimension. $D_i(\bm{\theta})$ is a function of the fit parameters that is higher for parameters that fluctuate more in the $i$th dimension. In this model we used third order finite differences to capture these fluctuations. For example, $D_{a'}$ is the sum of the squared third order finite differences between $\theta_{ijk}$ with the same $j$ and $k$. Third order differences were selected because for $n$-th order finite difference penalties, the first $n-1$ moments of the data will be unaffected by smoothing. Thus, for any values of $\lambda_i$ the mean and standard deviation of the distribution are left unchanged. See \cref{sec:smooth} for details on how values for $\lambda_i$ were chosen.

In order to have tractable fitting times we made use of general linear array modeling (GLAM) with sparse arrays. GLAM fitting speeds up the typical iterative GLM fitting procedure by orders of magnitude when working with high dimensional data. The data in this case is three dimensional, and so the array of three-dimensional basis splines that are fit to the data can be represented as the Kronecker product of three arrays of one-dimensional basis splines. By making use of properties of the Kronecker product GLAM lets us bypass the computation of particularly large matrices needed in the traditional fitting procedure. More detail on GLAM fitting can be found in ref.~\cite{10.1111/j.1467-9868.2006.00543.x}. 

Sampling from $f(a', b';E)$ can be performed via rejection sampling. An independent approach using iterative grid sampling was also developed, and is described in \cref{sec:sampling}. Both methods serve as cross checks of one another, and yield consistent results.

\subsection{Parametric model of the amplitude}
\label{sec:ltot}

In refs.~\cite{Raedel2012,Radel:2012ij} distributions of $\ltot$ from Geant4 were fit to a normal distribution at each energy. Two examples are shown in the right panel \cref{fig:example_showers} for \SI{1}{\tera \eV} $e^-$ and $\pi^+$ as dashed and dotted black lines, respectively. For comparison, the corresponding FLUKA $e^-$ ($\pi^+$) MC distribution is shown in blue (orange) and exhibits left (right) skew that deviates from the normal distribution. Additional nuclear effects which were not enabled in the prior Geant4 simulations but were in FLUKA lead to a noticeable tail in the $\ltot$ distribution of EM showers. The right panel of \cref{fig:emvspi} shows their energy dependence, highlighting that the skew is consistent across different energies.

In order to more accurately model FLUKA $\ltot$ distributions, and due to the differing skewness, new functional forms for EM and hadronic showers were used. The $\ltot$ distribution from hadron primaries are modeled by a skew normal distribution~\cite{Azzalini1999Statistical}, while $e^-$ and $\gamma$ primaries rely on the normal-inverse Gaussian (NIG) distribution~\cite{10.1111/1467-9469.00045}. Both are implemented in the SciPy toolkit~\cite{2020SciPy-NMeth}, from which their definitions are adapted. The skew normal is a three parameter distribution with shape, $\alpha$, location, $\xi$, and scale, $\omega$, defined in terms of $z \equiv (x - \xi)/\omega$ as
\begin{equation}
    \mathrm{SN}(x; \alpha, \xi, \omega) = \frac{2 \phi(z) \Phi(\alpha z)}{\omega},
    \label{eq:sn}
\end{equation}
where $\phi$ and $\Phi$ are the PDF and cumulative distribution function (CDF) of the standard normal distribution. The NIG is a four parameter distribution with two shape parameters $\alpha$ and $\beta$, as well as the aforementioned location and scale parameters. It can be similarly written in terms of $z$ as
\begin{equation}
    \mathrm{NIG}(x; \alpha, \beta, \xi, \omega) = \frac{\alpha K_1(\alpha \sqrt{1+z^2})}{\omega \pi \sqrt{1+z^2}} \exp(\sqrt{\alpha^2 - \beta^2} + \beta z),
    \label{eq:ng}
\end{equation}
with $\alpha > 0$, $|\beta| \leq \alpha$, and $K_1$ the modified Bessel function of the second kind.

\begin{figure}[tbp]
\includegraphics[width=0.5\linewidth]{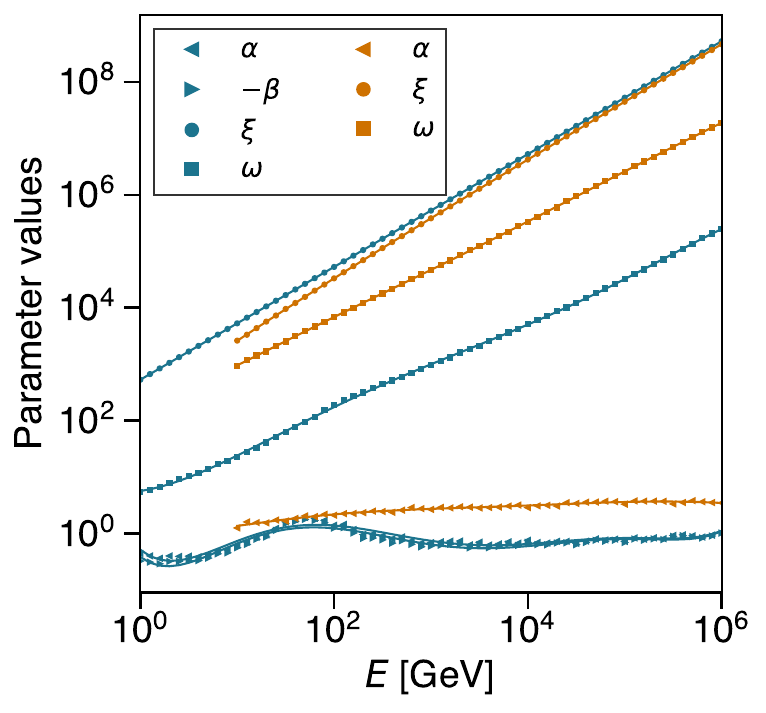}
\includegraphics[width=0.49\linewidth]{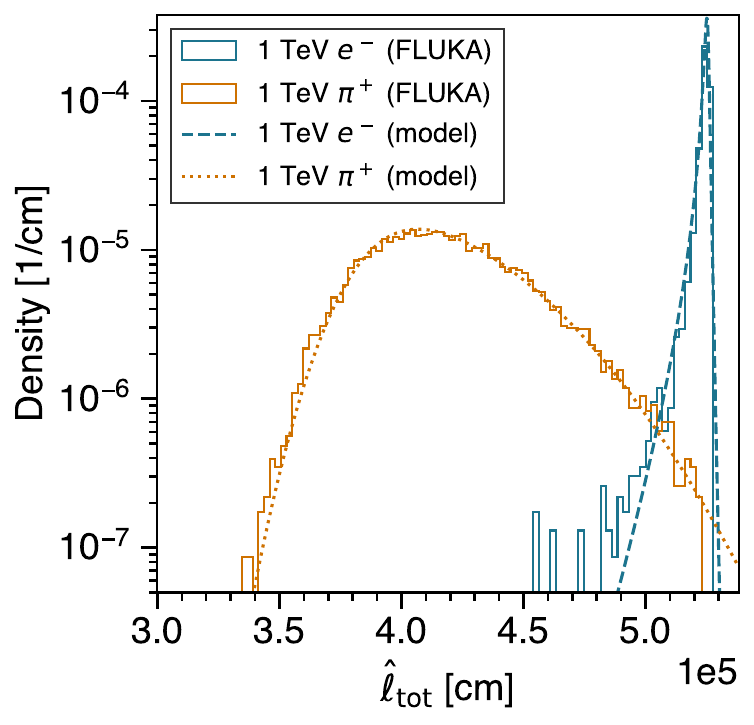}
\caption{The FLUKA $\ltot$ distributions in the right panel are identical to those in that of \cref{fig:example_showers}. A fit to a normal-inverse Gaussian (skew normal) distribution for $e^-$ ($\pi^+$) results in the probability density function indicated by the dashed blue (dotted orange) line. Both are improved descriptions of the data compared to the black dashed and dotted lines in \cref{fig:example_showers} from refs.~\cite{Raedel2012,Radel:2012ij}. The left panel shows fitted parameters as a function of the primary $e^-$ ($\pi^+$) energy in blue (orange) using different markers. For $e^-$ the parameters enter \cref{eq:ng}, and $\pi^+$ \cref{eq:sn}. Predictions from $g_p(E; \bm{\hat{t}}_p, s_p)$ are shown as corresponding lines. See text for more details.
\label{fig:ltot_fits}}
\end{figure}
At each of the primary energy steps, the parameters are fitted to the sample of $\ltot$ values from simulation via maximum likelihood. Once obtained for all 51 (61) hadron (EM) energies, the parameter values themselves are used to fit a set of polynomials in $\log_{10} E$ in order to model the parameters as a function of energy. More explicitly, we define
\begin{equation}
    g_p(E; \bm{t}_p, s_p) \equiv s_p \exp \left\{ \sum_{i=0}^{6} t_{p,i} (\log_{10} E)^i \right\},
    \label{eq:ltot_g_p}
\end{equation}
where the subscript $p$ indicates the parameter being modeled, $\bm{t}_p$ are coefficients being fitted and $s_p$ a sign term. Minimizing
\begin{equation}
    \chi^2(\bm{t}_p) = \sum_j (\ln s_p p_j - \ln s_p g_p(E_j; \bm{t}_p, s_p))^2,
    \label{eq:ltot_chi2}
\end{equation}
where the sum proceeds over simulated energy values, $E_j$, and $p_j$ corresponds to the parameter value obtained at that energy using either \cref{eq:sn} or \cref{eq:ng}, yields $g_p(E;\hat{\bm{t}}_p, s_p)$. Since the parameters values can be negative, the sign term, $s_p \in \{+1, -1\}$, allows the fit to proceed in terms of the natural logarithm. Besides $s_\beta=-1$, it is set to 1 for all other parameters.

To illustrate, the right panel of \cref{fig:ltot_fits} shows the NIG (SN) PDF in dashed blue (dotted orange) obtained from \cref{eq:ltot_g_p} for a \SI{1}{\tera \eV} $e^-$ ($\pi^+$). The underlying $\ltot$ distributions from FLUKA are shown as well, and are identical to that of the right panel of \cref{fig:example_showers}. Agreement between the PDFs and underlying distributions are noticeably improved. The left panel shows fitted parameter values at discrete energies using \cref{eq:ng} for $e^-$ (blue) and \cref{eq:sn} for $\pi^+$ (orange). Shape parameters are indicated by triangles, location by circles, and scale by squares. Lines that appear to interpolate the markers correspond to predictions from \cref{eq:ltot_g_p}. To reasonable approximation $\xi$ scales linearly with energy~\cite{Raedel2012,Radel:2012ij}. Deviations from linearity can be seen by comparing $\xi$ between the blue and orange, and in order to accurately model all parameters a sixth degree polynomial in $\log_{10} E$ is fitted according to \cref{eq:ltot_chi2}. This procedure is repeated for all particles listed in \cref{sec:flukaconf}.

\subsection{Accuracy}
\label{sec:accuracy}

\begin{figure}[tbp]
    \includegraphics[width=0.5\linewidth]{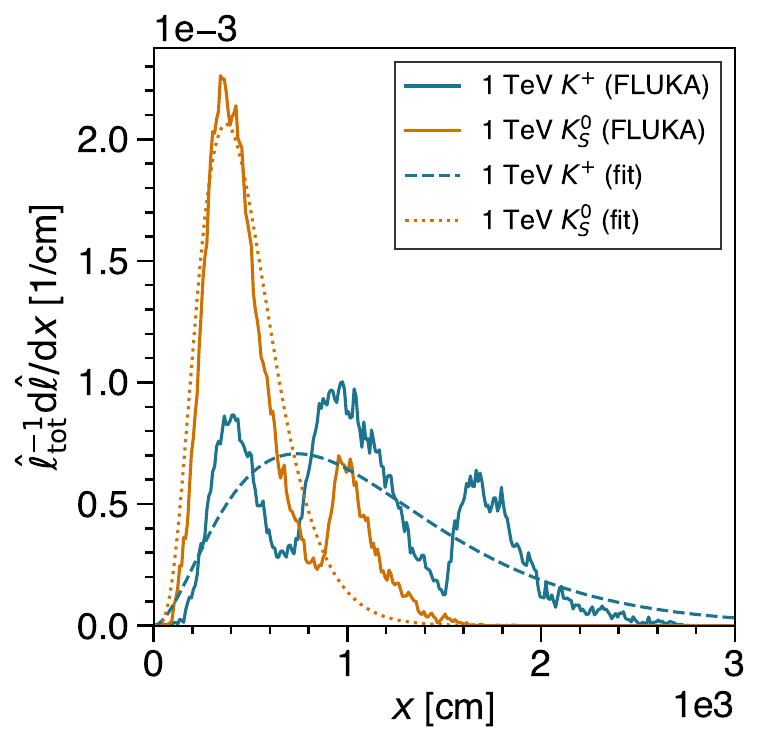}
    \includegraphics[width=0.49\linewidth]{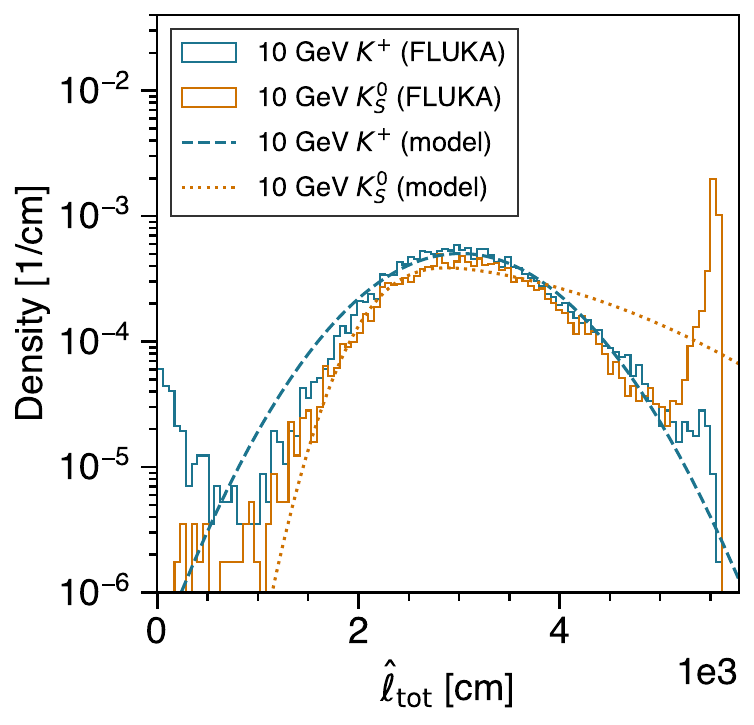}
    \caption{The left panel shows example outlier $\ltot^{-1} \dldx$ distributions from FLUKA, for which the gamma distribution (dashed and dotted lines) is a poor approximation. The blue (orange) solid line corresponds a FLUKA simulated \SI{1}{\tera \eV} $K^+$ ($K_S^0$). Both exhibit multiple peaks. The right panel shows the $\ltot$ distribution for a \SI{10}{\giga \eV} $K^+$ (blue) and $K^0_S$ (orange). The excess near zero for $K^+$ corresponds to decay channels with neutrinos. For $K_S^0$, a similar excess is visible near \SI{55}{\m}, corresponding to its decay to two $\pi^0$s, which immediately decay to photons. The skew normal distribution does not model these features that occur at lower energies.
    \label{fig:outliers}}
\end{figure}

To test accuracy, it is important to check how well gamma distributions represent the shape of particle showers, and if the models discussed in \cref{sec:shape} and \cref{sec:ltot} capture the fitted gamma and calculated $\ltot$ distributions from FLUKA simulations. Challenges can arise at lower energies, which exhibit higher fluctuations in shape, as suggested by the left panel of \cref{fig:statistics}. Even at higher energies, the stochastic nature of interaction processes can occasionally lead to multiple peaks in the shower profile, and the left panel of \cref{fig:outliers} shows examples for a \SI{1}{\tera \eV} $K^+$ primary in blue and $K_S^0$ in orange. The fitted gamma distribution to each is shown as dashed blue and dotted orange lines for $K^+$ and $K_S^0$, respectively, and while it is able to capture the dominant shape there is only a single mode.

For hadron primaries, excesses near the $\ltot$ boundaries can occur when the particle decays before first interaction. The right panel of \cref{fig:outliers} shows the $\ltot$ distribution from FLUKA as solid histograms, for \SI{10}{\giga \eV} $K^+$ in blue and $K_S^0$ in orange. At this energy, the primary kaons can decay before it interacts and as $K^+$ often decays to neutrinos, which escape undetected, an excess near $\ltot =0$ is visible. This low-$\ltot$ tail is what motivated a \SI{0.5}{\%} quantile threshold, mentioned earlier in this \namecref{sec:model}. In contrast, $K_S^0 \rightarrow \pi^0 + \pi^0$ at \SI{30.69}{\%}. The $\pi^0$s immediately decay to photons and appear as an EM shower, leading to the peak near $\ltot = \SI{55}{\m}$. The fitted skew normal distributions, dashed blue for $K^+$ and dotted orange for $K_S^0$ does not capture these features that can occur at lower energies, and are somewhat pulled by them. At higher energies, interaction of the primary overcomes decay and the $\ltot$ distribution becomes similar to that shown in the right panel of \cref{fig:ltot_fits}.

\begin{figure}[tbp]
    \centering
    \includegraphics[width=\linewidth]{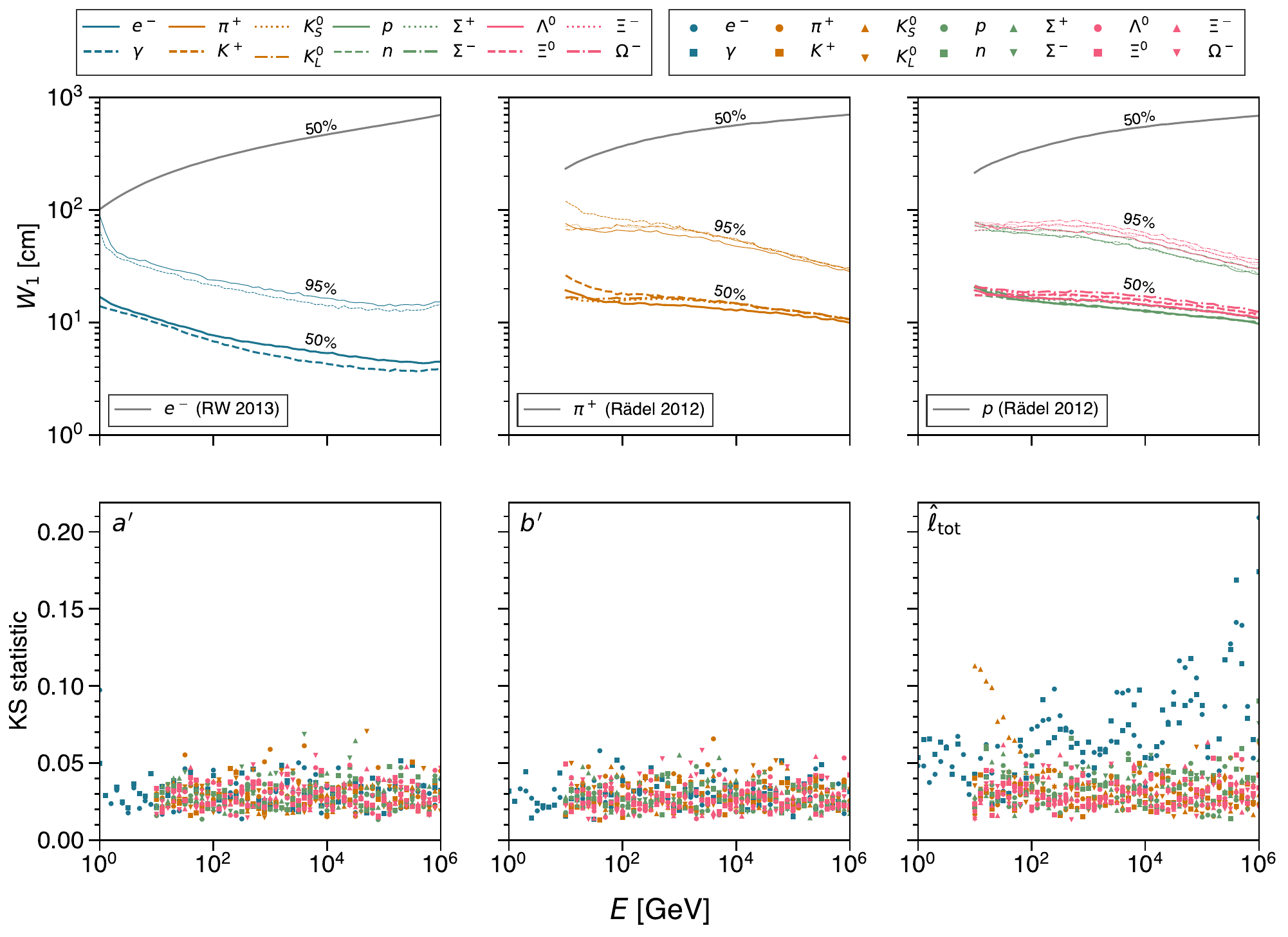}
    \caption{The colored lines in the top three panels shows \SI{50}{\%} and \SI{95}{\%} quantiles of the Wasserstein distance, $W_1$, as a function of energy for all primaries. Additionally, the median obtained assuming the average parametrizations from refs.~\cite{Raedel2012,Radel:2012ij} are included as gray lines. From left to right, the panels correspond to EM, mesonic, and baryonic primaries, as indicated by the top-left legend. The bottom three panels show the KS statistic for $a'$, $b'$ and $\ltot$ also as a function of energy for all primaries. The different primaries are shown as different markers, listed in the top-right legend. For each primary, at a given energy, the KS statistic was computed by comparing \si{1000} samples from the model against the values calculated directly from the \si{10000} FLUKA simulations.}
    \label{fig:accuracy}
\end{figure}

The gamma distribution was successfully fitted to almost every simulated shower. At most a single fit failed out of \si{10000} simulations at a given energy, mostly below \SI{1}{\tera \eV}, and no more than four over all energies for any given primary. To quantify the accuracy, every gamma distribution was compared against its FLUKA simulation and the Wasserstein distance, $W_1 = \int_{-\infty}^\infty |\tilde{F}(x) - F(x)|dx$, was computed~\cite{10.1561/2200000073}. Here, $\tilde{F}(x)$ is the sum of the binned FLUKA $\ltot^{-1} \dldx$ up to distance $x$, and $F(x)$ the CDF of its fitted gamma distribution. Note that each simulation thus corresponds to an independent data point. For each simulated energy, the resulting $W_1$ \SI{50}{\%} and \SI{95}{\%} quantiles are shown as colored lines in the top three panels of \cref{fig:accuracy} for all primaries, listed in the top-left legend. For comparison, the median obtained assuming the average-shape parametrizations from refs.~\cite{Raedel2012,Radel:2012ij} are included as gray lines. A significantly improved characterization of the shape is observed with the individually fitted gamma distributions. The metric $W_1$ was chosen, as opposed to a statistic, since the intent is not to claim that simulated shower shapes correspond to the gamma distribution; as discussed earlier, the gamma distribution is a simple approximation. Instead, $W_1$ is a measure of proper distance between two distributions, and highlights the energy dependence and improved agreement relative to refs.~\cite{Raedel2012,Radel:2012ij}.

Finally, to check model accuracy, a two sample Kolmogorov-Smirnov (KS) test was performed. For each primary, at each simulated energy, \si{1000} samples of $(a', b')$ from $f(a', b'; E)$ and $\ltot$ using either \cref{eq:sn} (hadrons) or \cref{eq:ng} (EM) based on parameter predictions from \cref{eq:ltot_g_p} were drawn. The KS statistic was computed by comparing those samples, marginal for $a'$ and $b'$, against fitted $(a', b')$ or calculated $\ltot$ values for the \si{10000} FLUKA simulations. The results are shown in the lower three panels of \cref{fig:accuracy} for all simulated energies, with primaries indicated by different markers given in the top-right legend. The statistic is the maximum difference between the empirical cumulative distributions of both samples, and was chosen to highlight that, in all cases of $a'$ and $b'$ and most cases of $\ltot$, the sampled parameters are consistent with those obtained from simulation. Larger differences in $\ltot$ occur for $K_S^0$ at energies below \SI{100}{\giga \eV} due to its decays, shown in the right panel of \cref{fig:outliers}, as well as $e^-$ and $\gamma$, which are caused by slight inaccuracies in parameter estimation with $g_p(E; \bm{\hat{t}}_p, s_p)$ and visible in the left panel of \cref{fig:ltot_fits}. The latter was tested to be the case by sampling from the fitted $\ltot$ distribution for each simulation directly, and computing the corresponding KS statistic.

\section{Application to high-energy neutrino interactions}
\label{sec:application}

The models described in \cref{sec:model} are for a single primary particle. In collisions where multiple final state particles are produced, the model can be applied individually to each one. Here, we use PYTHIA8~\cite{Sjostrand:2014zea} to simulate electron-neutrino DIS interactions off a proton target. Final state particles with a proper lifetime of $c\tau_0 > \SI{0.5}{\mm}$ are stored directly. Particles with $c\tau_0 < \SI{0.5}{\mm}$, which includes most short-lived particles like $D$-mesons and $B$-mesons, are propagated and decayed by PYTHIA. Longer-lived decay products are then stored. For heavy mesons, at energies above a few tens of TeV the interplay of decay and interaction lengths makes their parametrization challenging, and a MC approach is necessary for accurate simulation. Further assuming that antiparticles can use the same parametrizations as their particle counterparts, the final list of particles and their energies can be used to individually sample $\dldx$.

\begin{figure}[tbp]
    \centering
    \includegraphics[width=\linewidth]{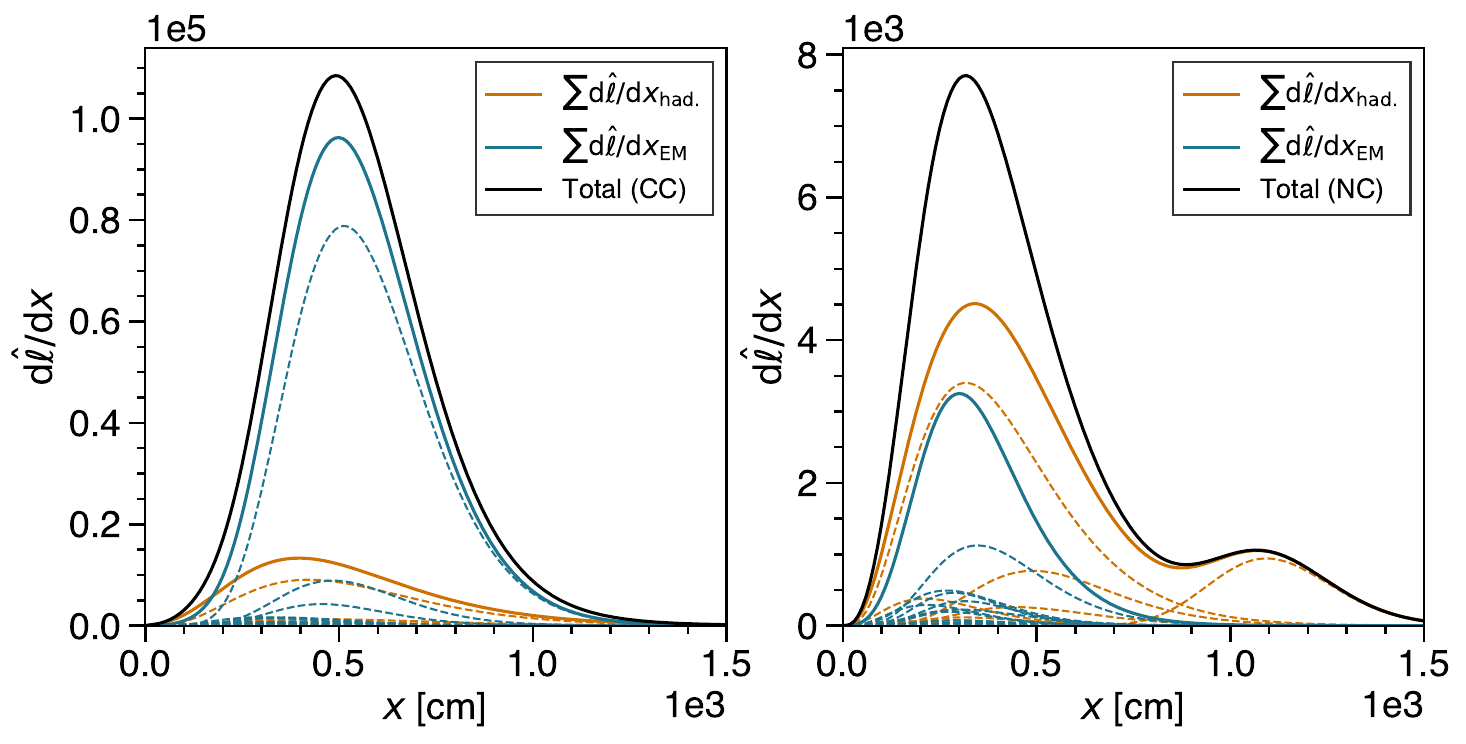}
    \caption{Two example $\dldx$ distributions from our model for a \SI{100}{\tera \eV} electron-neutrino interaction in ice. The left panel corresponds to a charged current (CC) interaction, and the right panel shows a neutral current (NC) interaction. Blue (orange) dashed lines show sampled $\dldx$ distributions for EM (hadron) primaries, and solid lines show the sum for each category. The total is shown in black. Note that since $\pi^0 \rightarrow \gamma + \gamma$, there is an EM component for the NC case.}
    \label{fig:pythia}
\end{figure}
\Cref{fig:pythia} shows the $\dldx$ distributions for a \SI{100}{\tera \eV} electron-neutrino DIS interactions. Note that, as described above, nuclear effects are ignored and short-lived particles are allowed to decay at some distance away from the interaction vertex. The left panel corresponds to a charged-current (CC) interaction and the right panel a neutral current (NC) interaction. In NC interactions, only a fraction of the neutrino energy remains visible, depending on the inelasticity. Blue (orange) dashed lines show the sampled $\dldx$ distributions, consisting of both a sampled $\ltot$ and a sampled gamma distribution, for EM (hadron) primaries, where EM is taken as $e^-$ or $\gamma$. Solid lines show the sum of the two categories, and the total is shown in black. An EM component is visible in the NC case shown in the right panel due to $\pi^0 \rightarrow \gamma+\gamma$. Notably, a fluctuation for one of the hadrons causes a second peak in the $\dldx$ in the right panel, although CC interactions that typically contain a single high-energy electron will be more similar to the profile shown in the left panel.

A more comprehensive visualization is presented in \cref{fig:pythia-80} in \cref{sec:appendix-pythia}. There, \si{80} $\dldx$ outcomes are shown, with the first two corresponding to that of \cref{fig:pythia}. While NC interactions (blue panels) tend to present more structure at lower amplitudes, CC interactions exhibit variation in terms of the overall extension and, in a few cases, additional structure is visible.

\begin{figure}[tbp]
    \centering
    \includegraphics[width=\linewidth]{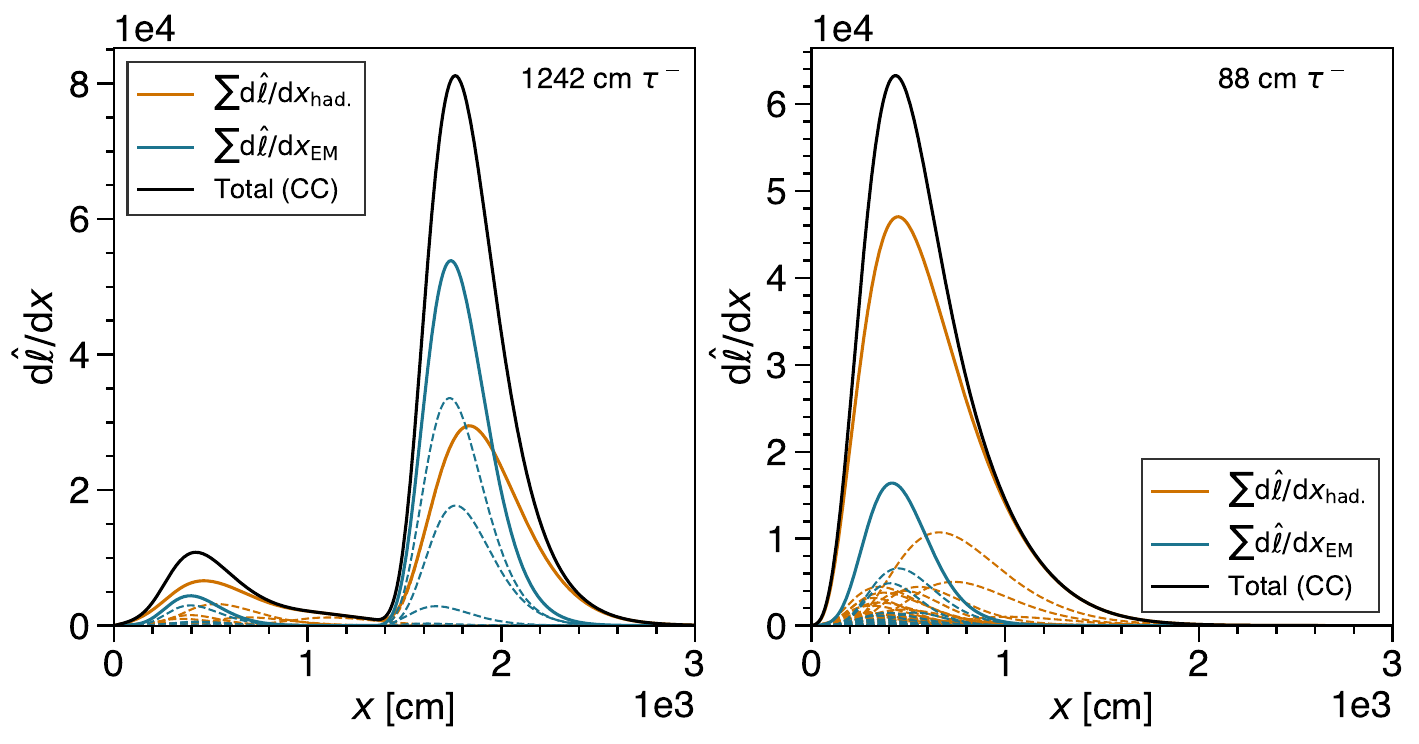}
    \caption{Same as \cref{fig:pythia} but for \SI{100}{\tera \eV} tau-neutrino CCDIS in ice and x-axis extended up to \SI{30}{\m}. In the left (right) panel, the outgoing $\tau^-$ decays to two $\pi^0$ ($\pi^-$) and a $\pi^-$ ($\pi^+$), in addition to a $\nu_\tau$. The hadrons, or their immediate decay products, are input to the $\dldx$ model. The first peak in the left panel is due to hadrons produced in the initial DIS interaction. The $\tau^-$ decay lengths are indicated in the panels.}
    \label{fig:pythia-tau}
\end{figure}

The examples discussed so far focuses on electron-neutrino DIS, but the shower parametrizations are applicable to other processes as well. High-energy muons typically lose energy stochastically via Bremsstrahlung, pair production or photonuclear interactions, all of which result in primaries that initiate particle showers. Decays of tau leptons or $W$ bosons can also produce such primaries. To illustrate, \cref{fig:pythia-tau} shows the $\dldx$ distribution for two \SI{100}{\tera \eV} tau-neutrino CCDIS interactions, where both the initial interaction and subsequent tau decay are simulated with PYTHIA8. The left (right) panel corresponds to a scenario where the outgoing $\tau^-$ travels \SI{1242}{\cm} (\SI{88}{\cm}) and then decays to two $\pi^0$ ($\pi^-$) and a $\pi^-$ ($\pi^+$), in addition to a $\nu_\tau$. The hadrons, or their immediate decay products in the case of $\pi^0$, are inputs for the model, from which $\dldx$ distributions are then sampled. In the left panel, both the initial DIS and tau decay are visible as two distinct peaks in the $\dldx$ distribution. Note that stochastic energy losses along the tau track are not simulated, but should be a relatively small contribution at energies below \SI{100}{\tera \eV}. The simulation of such stochastic processes for taus and muons can be performed using software such as PROPOSAL~\cite{Koehne:2013gpa}.

More generally, for any high-energy simulation interested in the Cherenkov light yield of particle showers, physics processes can be tracked using MC up to a certain threshold. Beyond that, a switch to using the parametrizations detailed in \cref{sec:model} can yield a substantial speed-up in the simulation. The inverse problem of reconstruction can also benefit from a more accurate modeling of Cherenkov emission. In particular, elongation or fluctuations of the shower profile can improve directional reconstruction, if the reconstruction algorithm is flexible enough to fit a superposition of point-like showers~\cite{hallen2013measurement,IceCube:2024csv}. Such an approach may also result in more accurate energy reconstructions, by fitting the individual sub-shower energies. Further, machine learning models trained on simulation would benefit from increased simulation fidelity. Using modern tools, it may be possible to distinguish, on a statistical basis, showers which are more EM-dominated, as seen in \cref{fig:pythia-80}.

\section{Discussion}
\label{sec:discuss}

Using sophisticated particle simulation software like FLUKA~\cite{Ferrari:2005zk,Ballarini:2024isa}, the most subtle details of particle interactions can be modeled to high accuracy. In applications to high-energy Cherenkov calorimeters, such as neutrino telescopes, it is essential to model detectable signatures that are relevant for analyses. While existing models have approximated the energy deposition characteristics in particle showers, the method outlined in this paper more accurately captures the shape and scale of the Cherenkov yield. A deterministic model cannot represent variations in the position and dispersion of a shower~\cite{Radel:2012ij,Raedel2012}. With the addition of relatively minor computational complexity we can capture both of these factors, as detailed in \cref{sec:shape} and \cref{sec:ltot}.

The datasets used in this study consist of particle showers from a number of the most common primaries, described in \cref{sec:flukaconf}. Although antiparticles were not simulated, to good approximation they should be similar to their particle counterparts. If desired, one need only run FLUKA simulations to sufficiently high statistics for antiparticles, and fit the model as described above. Short-lived hadrons tend to decay before their first interaction, and their longer-lived decay products should fall into the list of available particles. The exception to this in our energy range of interest are the heavy mesons, for which decay-interaction interplay carries additional complexities that can be modeled more accurately with MC than the parametrizations described in \cref{sec:model}.

A simple-to-use Python package containing all the models is available~\cite{acem}. The best-fit B-spline coefficients, $\bm{\hat{\theta}}$, their defining knots, as well as the polynomial coefficients, $\bm{\hat{t}}_p$, for modeling $\ltot$ are saved in the NumPy format~\cite{harris2020array}. As these are simply arrays of floats, they can be ported to other frameworks and languages in a relatively straightforward manner using existing B-spline libraries~\cite{Whitehorn:2013nh}.

Several simplifications were assumed in the development of this model, which future work can improve upon. The most obvious of these is that only a one-dimensional $\dldx$ is modeled, and a fully three-dimensional model likely would require new methods. Even in one dimension, the gamma distribution is an approximation and does not capture multiple peaks, which may arise due to a combination of multiple sub-shower components. A couple examples of outliers are shown in the left panel of \cref{fig:outliers}. The factorization of shape parameters $(a', b')$ from the amplitude $\ltot$ also neglects any shape-amplitude correlations. This may be a reasonable assumption for most of the energy ranges studied here, but breaks down in cases such as decay. Examples of $\ltot$ distributions at lower energies, where decay begins to dominate, are shown in the right panel of \cref{fig:outliers}. Finally, muons from hadronic decays were not included as discussed in \cref{sec:flukaconf}. Typically, these muons do not carry away large fractions of the primary energy, and can be handled separately, but could be included in a more complete model.

Existing simulation of high-energy neutrino interactions at neutrino telescopes such as IceCube does not incorporate shower-to-shower fluctuations, and instead rely on the average profiles from refs.~\cite{Radel:2012ij,Raedel2012}. When such fluctuations are included, they can lead to shower profiles that differ substantially from the mean, and can modify the Cherenkov light yield of neutrino interactions as shown in \cref{sec:application}. Individual showers that peak at upwards of \SI{10}{\m} from the interaction vertex are possible. For analyses searching for rare signals based on small separation distances, it becomes more important to accurately simulate in-medium particle showers. The models presented in \cref{sec:model} are a step towards more accurate representations of the Cherenkov yield from high-energy particle showers in ice and water, and may be of use in simulations for existing and upcoming experiments.

\acknowledgments

We thank A. Ferrari for helpful advice regarding FLUKA usage, M. Jin for insights on charm hadrons and D. Chirkin for useful discussions. TY, EY and LL were supported in part by NSF grants PHY-2513077 and IIS-2435532, and by the University of Wisconsin Research Committee with funds granted by the Wisconsin Alumni Research Foundation. AF acknowledges support from Academia Sinica grant AS-GCS-113-M04 and the National Science and Technology Council grant 113-2112-M-001-060-MY3. TY gratefully acknowledges the Center for High Throughput Computing (CHTC) at the University of Wisconsin–Madison for enabling high-energy FLUKA simulations to be performed at scale.


\appendix
\section{Additional enabled physics in FLUKA}
\label{sec:appendix-fluka}
Additional physics settings enabled in the FLUKA simulations described in \cref{sec:flukaconf} are given below. For a description of their effects see ref.~\cite{Ballarini:2024isa} and the FLUKA online manual at \url{https://www.fluka.eu}.
{\small
\begin{verbatim}
PHYSICS          3.0                                                  EVAPORAT
PHYSICS          1.0                                                  COALESCE
PHYSICS           1.     0.005      0.15       2.0       2.0        2.IONSPLIT
PHOTONUC         1.0                         3.0    @LASTMAT
PHOTONUC         1.0                         3.0    @LASTMAT          ELECTNUC
IONTRANS    HEAVYION
\end{verbatim}
}

\section{Smoothing parameters}
\label{sec:smooth}
The values for the smoothing parameters, $\lambda_i$ (for each dimension $i$) were found by minimizing the Akaike information criterion (AIC), defined as
\begin{equation}
    \label{eq:aic}
    \text{AIC} = \text{dev}(y,\bm{\theta}) + 2\text{dim}(\bm{\theta}).
\end{equation}
Here, $\text{dev}$ is the deviance, a metric of the closeness between the data and the model, and $\text{dim}$ is the effective dimension of the model space, or the degrees of freedom we had available in fitting. This is calculated as the trace of the hat-matrix for the general linear model~\cite{10.1214/ss/1038425655}.

Given that the AIC can only be calculated for a fitted model, and fitting is a computationally intensive process, we cannot verify that the smoothing parameters are optimal. They were found through trial and error, testing different sets of smoothing parameters until deviations in any direction yielded a higher AIC value. The minimum AIC achieved corresponded to $\lambda_{a'} = \lambda_{b'}= 0.1$ and $\lambda_E = 0.01$.

The effect of smoothing can be quantified by the effective dimension. An unsmoothed model would have complete freedom to adjust $c_{a'}c_{b'}c_E$ basis splines, leading to an effective dimension of 2312. The presence of smoothing puts constraints on the array, $\bm{\theta}$, decreasing the model's degrees of freedom. For the smoothing used we calculated the effective dimension to be about 288.

\section{Iterative grid sampling}
\label{sec:sampling}
An independent, iterative grid sampling routine was developed to complement the usual rejection sampling from $f(a', b';E)$. Given a particular energy we integrate $f$ in each square region in the two-dimensional grid of spline regions. By calculating the cumulative distribution of these regions and sampling a number from a uniform distribution, we select which spline region to work in. With a single spline region selected, the probability distribution is represented not as an exponentiated piece-wise polynomial, but simply a single exponentiated polynomial. We then divide the region in halves, and integrate each half. We sample another uniformly distributed number and select a half to continue from. We alternate divisions in the $a'$ and $b'$ dimensions, and repeat this process until the desired precision is reached. Each region has dimensions $1/c_{a'}$  by $1/c_{b'}$. If we perform $N$ divisions in each dimension we can get a precision of $2^{-N}/c_{a'}$ and $2^{-N}/c_{b'}$ in our values for $a'$ and $b'$ respectively.

\section{Additional results for \SI{100}{\tera \eV} electron neutrino DIS}
\label{sec:appendix-pythia}
\begin{figure}[tbp]
\includegraphics[width=\linewidth]{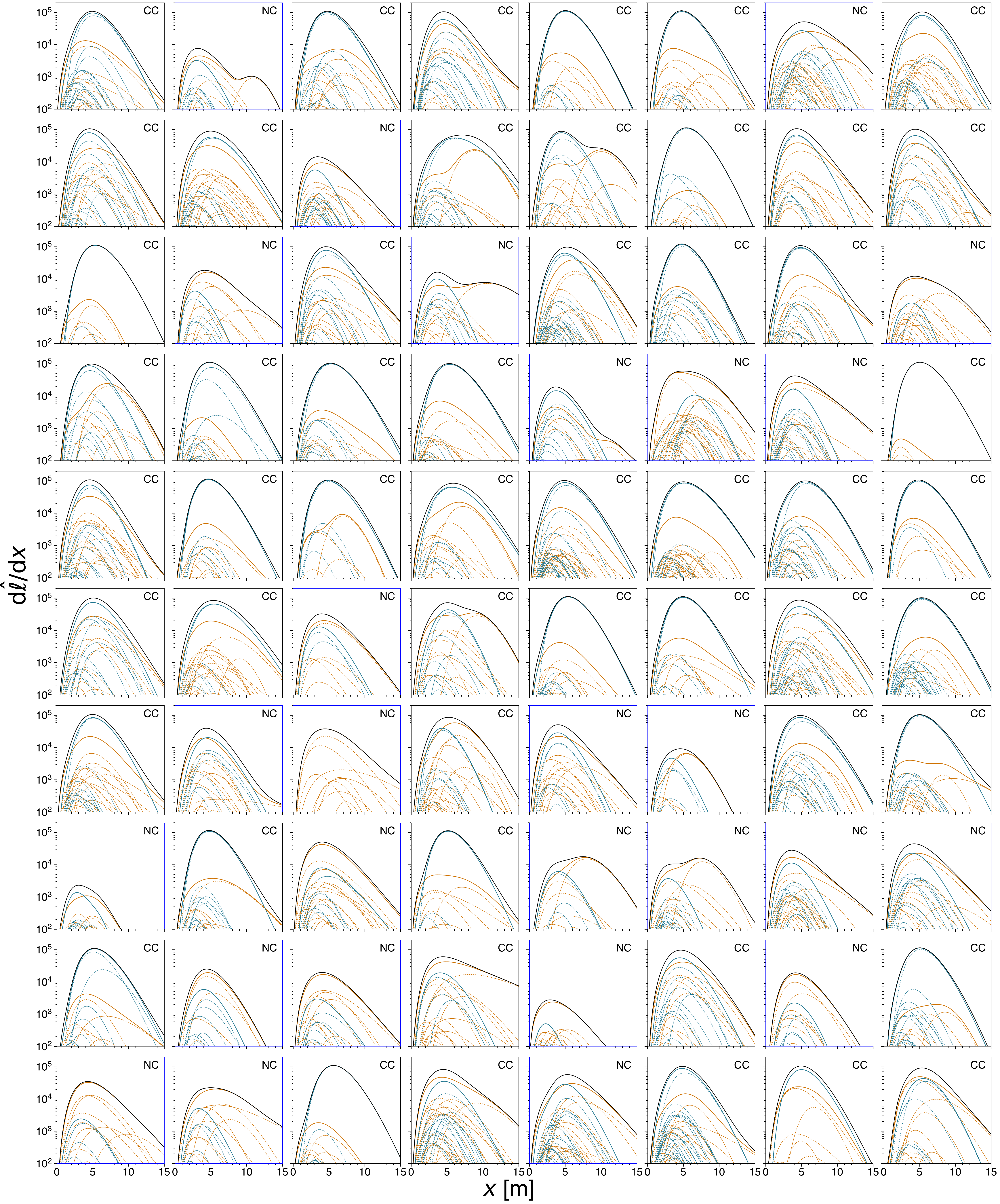}
\caption{Eighty $\dldx$ results based on \SI{100}{\tera \eV} electron-neutrino interactions generated with PYTHIA8~\cite{Sjostrand:2014zea}. The first two correspond to that of \cref{fig:pythia}. Panels highlighted in blue correspond to NC interactions, otherwise CC. Within each panel, blue (orange) dashed lines show sampled $\dldx$ distributions for EM (hadron) primaries, and solid lines show the sum for each category. The total is shown in black. To give a sense of both the scale and shape a logarithmic scale is used in the y-axis.
\label{fig:pythia-80}}
\end{figure}

\Cref{fig:pythia-80} shows \si{80} $\dldx$ profiles of \SI{100}{\tera \eV} electron-neutrino interactions, as discussed in \cref{sec:application}. The first two correspond to that shown in \cref{fig:pythia}. Each neutrino interacts off a fixed proton target, with hadronization and decays of particles up to $c\tau_0<\SI{0.5}{\mm}$ simulated with PYTHIA8~\cite{Sjostrand:2014zea}. The resulting final state $e^-$, $\gamma$ and hadrons are passed to the model of \cref{sec:model}, which is used to sample $\dldx$ according to their energies. In \cref{fig:pythia-80}, panels highlighted in blue correspond to NC interactions, otherwise CC.

\clearpage


\bibliographystyle{JHEP}
\bibliography{biblio.bib}






\end{document}